\documentclass[superscriptaddress,prl,floatfix,twocolumn,amsmath,amssymb,aps]{revtex4-1}

\usepackage{graphicx}
\usepackage{amsmath}
\usepackage{amssymb}
\usepackage{dcolumn}
\usepackage{color}
\usepackage{multirow}
\usepackage{bm}
\usepackage{subfigure}
\usepackage{booktabs}

\begin{document}
	\title{Phonon-mediated exciton relaxation in two-dimensional semiconductors: selection rules and relaxation pathways}
	\author{Xiao-Wei Zhang}
	\affiliation{International Center for Quantum Materials and School of Physics, Peking University, Beijing, China}
	\affiliation{Department of Materials Science and Engineering, University of Washington, Seattle, WA 98195, USA}
	\author{Kaichen Xie}
	\affiliation{Department of Materials Science and Engineering, University of Washington, Seattle, WA 98195, USA}
	\author{En-Ge Wang}
	\email{egwang@pku.edu.cn}
	\affiliation{International Center for Quantum Materials and School of Physics, Peking University, Beijing, China}
	\affiliation{Ceramic Division, Songshan Lake Lab, Institute of Physics, Chinese Academy of Sciences, Guangdong, China}
	\affiliation{School of Physics, Liaoning University, Shenyang, China}
	\author{Ting Cao}
	\email{tingcao@uw.edu}
	\affiliation{Department of Materials Science and Engineering, University of Washington, Seattle, WA 98195, USA}
	\author{Xin-Zheng Li}
	\email{xzli@pku.edu.cn}
	\affiliation{Interdisciplinary Institute of Light-Element Quantum Materials and Research Center for Light-Element Advanced Materials, State Key Laboratory for Artificial Microstructure and Mesoscopic Physics, Frontier Science Center for Nano-optoelectronics and School of Physics, Peking University, Beijing 100871, China}
    \affiliation{Peking University Yangtze Delta Institute of Optoelectronics, Nantong, Jiangsu 226010, China}
	\date{\today}
\begin{abstract}
Exciton-phonon coupling (ExPC) is crucial for energy relaxation in semiconductors, yet the first-principles calculation of such coupling remains challenging, especially for low-dimensional
systems.
Here, an accurate algorithm for calculating ExPC is developed and applied in exciton relaxation
problems in monolayer WS$\text{e}_{2}$.
Considering the interplay between the exciton wave functions and electron-phonon coupling (EPC) matrix elements, we find that ExPC shows distinct selection rules from the ones of EPC.
By employing the Wannier exciton model, we generalize these selection rules, which state that the angular quantum numbers of the exciton must match the winding numbers of the EPC matrix elements for the ExPC to be allowed.
To verify our theory and algorithm, we calculate inter-valley exciton relaxation pathways, which agrees well with a recent experiment.
\end{abstract}
	
\maketitle
Phonons and their interactions with electronic quasiparticles play critical roles in the optical properties and relaxation dynamics of materials.
In conventional semiconductors such as GaAs, hot carrier thermalization has been mostly studied in the context of electron-phonon coupling.
In recent years, 2D materials have emerged as a fertile platform for studies
of novel optical properties absent in conventional material systems~\cite{song2013two,xiao2017excitons,manzeli20172d,wang2018colloquium,xu2014spin,mak2018light}.
In 2D semiconductors such as monolayer transition metal dichalcogenides, the excitons have large binding energies reaching a fraction of eV due to the reduced dielectric screening and quantum confinement effects~\cite{cardona2005fundamentals,wang2018colloquium}.
As a result, optical excitation can directly create sub-bandgap and tightly-bound excitons.
Phonon-mediated scattering between different excitonic states therefore becomes ubiquitous in the energy relaxation dynamics, exciton recombination, and photoluminescence of these materials~\cite{cardona2005fundamentals}.
Despite the serge of research interest in these exciton-phonon phenomena, most first-principles calculations and theoretical analyses so far have been based on electron-phonon coupling at one-electron level, with excitonic effects included either \textit{ad hoc} or only perturbatively.
At the one-electron level, group theory analysis based purely on the electron-phonon coupling (EPC) has been a convention~\cite{song2013transport,li2019emerging,liu2019valley,he2020valley}.
The role of excitonic interactions beyond the one-electron level, and how they affect the phonon scattering process, have been mostly neglected.
As such, it remains an open question whether exciton-phonon coupling (ExPC) has fundamentally different physical behaviors from EPC in these 2D systems.
In this letter, we take monolayer WSe$_{2}$, a multivalley 2D semiconductor, as an example to investigate exciton-phonon interactions and exciton relaxation pathways in atomically thin semiconductors.
The recently determined indirect-gap nature of this system implies interesting inter-valley exciton-phonon scattering~\cite{zhang2015probing,hsu2017evidence,elliott2020surface}.
Here, we go beyond the conventional group theory analysis that was based on the one-electron EPC matrix elements, and elucidate the connections between the simplified electron- or hole-phonon scattering problem and the exciton-phonon interaction problem.
We find that the ExPC resulted in distinct scattering pathways and completely different selection rules from EPC.
With excitonic effects included, our analyses show that the interplay between the exciton envelope functions on a two-particle basis and the EPC matrix elements plays crucial role in correctly describing ExPC.
For Wannier excitons at 2D, we discover that the change of angular momentum quantum number of the exciton and the winding number of EPC matrix elements must compensate each other such that ExPC can be allowed.
To verify our theory and algorithm, we calculate inter-valley exciton-phonon scattering times, which agrees well with a recent experiment.

General formulation for ExPC can be traced back to the mid of the last century~\cite{toyozawa1958theory,knox1963}.
The Hamiltonian of the exciton-phonon coupled system reads~\cite{knox1963}:
\begin{align}	H=&\sum_{S\bm{Q}}\Omega_{S\bm{Q}}a_{S\bm{Q}}^{\dagger}a_{S\bm{Q}}+\sum_{\bm{q}\nu}\left(b_{\bm{q}\nu}^{\dagger}b_{\bm{q}\nu}+\frac{1}{2}\right)\hbar\omega_{\bm{q}\nu} \nonumber \\
	&+ \sum_{SS^{\prime}\bm{Q}}\sum_{\bm{q}\nu}G_{S^{\prime}S}\left(\bm{Q},\bm{q}\nu\right)a^{\dagger}_{S^{\prime}\bm{Q}+\bm{q}}a_{S\bm{Q}}\left(b_{\bm{q}\nu}+b_{-\bm{q}\nu}^{\dagger}\right).
	\label{ex-ph-H}
\end{align}
In this Hamiltonian, the first term treats excitons as non-interacting bosons with the use of exciton creation and annihilation operators, $a_{S\bm{Q}}^{\dagger}$ and $a_{S\bm{Q}}$. $\Omega_{S\bm{Q}}$ is the exciton eigenvalue with the center-of-mass momentum $\bm{Q}$ and band index $S$.
$\Omega_{S\bm{Q}}$ can be solved from the Bethe-Salpeter equation (BSE)~\cite{strinati1988application, rohlfing2000electron}.
The eigenfunction (called the envelope function) of the BSE gives the exciton wave function,
\begin{equation}
 \chi_{S}^{\bm{Q}}(\bm{r}_{e},\bm{r}_{h})=\sum_{vc\bm{k}}A_{v\bm{k},c\bm{k}+\bm{Q}}^{S}\psi_{c\bm{k}+\bm{Q}}(\bm{r}_{e})\phi_{v\bm{k}}^{\ast}(\bm{r}_{h}).
\label{Ex-wfn}
\end{equation}
The exciton wavefunction consists of an envelope function $A_{v\bm{k},c\bm{k}+\bm{Q}}^{S}$ that carries the angular momentum of the exciton, and the Bloch wavefunction of the electron (hole) is $\psi_{c\bm{k}+\bm{Q}}$ ($\phi_{v\bm{k}}$).
The second term of Eq.~\ref{ex-ph-H} represents the non-interacting phonon system in harmonic approximation, where $b_{\bm{q}\nu}$ ($b_{\bm{q}\nu}^{\dagger}$) is the phonon annihilation (creation) operator and $\omega_{\bm{q}\nu}$ is the frequency.
The third term of Eq.~\ref{ex-ph-H} is the first-order ExPC Hamiltonian and the ExPC matrix element is defined as~\cite{knox1963}:
\begin{align}
	 G_{S^{\prime}S}(\bm{Q},\bm{q}\nu) &= \sum_{\bm{k}vcc^{\prime}}    A_{v\bm{k},c^{\prime}\bm{k}+\bm{Q}+\bm{q}}^{S^{\prime}\ast}A_{v\bm{k},c\bm{k}+\bm{Q}}^{S}g_{c^{\prime}c}(\bm{k}+\bm{Q},\bm{q}\nu)  \nonumber \\
	 &-\sum_{\bm{k}cvv^{\prime}}    A_{v\bm{k}-\bm{q},c\bm{k}+\bm{Q}}^{S^{\prime}\ast}A_{v^{\prime}\bm{k},c\bm{k}+\bm{Q}}^{S}g_{v^{\prime}v}(\bm{k}-\bm{q},\bm{q}\nu).
\label{ex-ph-G}
\end{align}
Here, $g_{c^{\prime}c}=\langle c^{\prime}\bm{k}+\bm{Q}+\bm{q}|\delta_{\bm{q}\nu}V^{\text{scf}}|c\bm{k}+\bm{Q}\rangle$ and $g_{v^{\prime}v}=\langle v^{\prime}\bm{k}|\delta_{\bm{q}\nu}V^{\text{scf}}|v\bm{k}-\bm{q}\rangle$ are EPC matrix elements.
Eq.~\ref{ex-ph-G} defines the relationship between the ExPC and EPC matrix elements, $G$ and $g$, and serves as the starting point of our analyses.
The minus sign ahead of the second term of Eq.~\ref{ex-ph-G} arises from the energy difference between the electron and hole, which contributes to the exciton energy.
\begin{figure}[b]
	\includegraphics[width=\columnwidth]{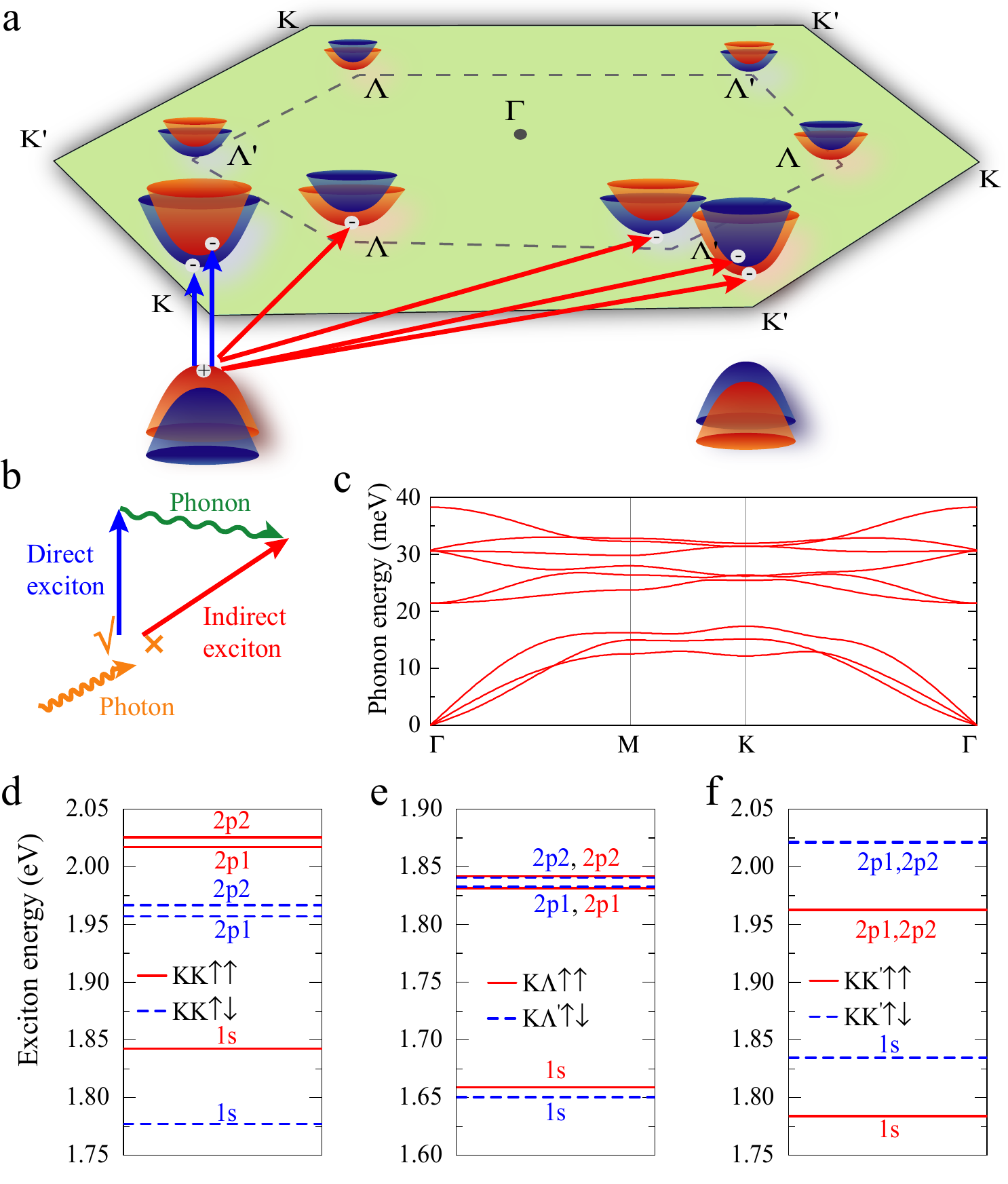}
	\caption{Excitons, phonons, and schematic exciton-phonon interactions in monolayer WSe$_2$. (a) The first BZ of monolayer WSe$_{2}$. The orange and blue parabolas denote the spin-up and spin-down bands, respectively. Inside the 1st BZ, three $\Lambda$ points and three $\Lambda^{\prime}$ points are connected by a dashed hexagon. On the boundary of the 1st BZ, three equivalent K points and three equivalent K$^{\prime}$ points are connected by a solid hexagon.
	(b) The schematic diagram of exciton-phonon scattering. A photon (orange wavy arrow) can create direct excitons (blue arrow) but can't create indirect excitons (red arrow) due to momentum mismatch.
	Phonon (green wavy arrow) can scatter direct excitons to indirect excitons by contributing a finite momentum.
	(c) Phonon dispersion along high-symmetry lines.
	(d)-(f) Eigenvalues of KK (d), K$\Lambda$ (e), K$\Lambda^{\prime}$ (e), and KK$^{\prime}$ (f) excitons, including 1$s$ and 2$p$, like-spin (solid red lines) and unlike-spin (dashed blue lines) states.
	}
	\label{fig1}
\end{figure}

\begin{figure}[b]
	\centering
	\includegraphics[width=1.0\columnwidth]{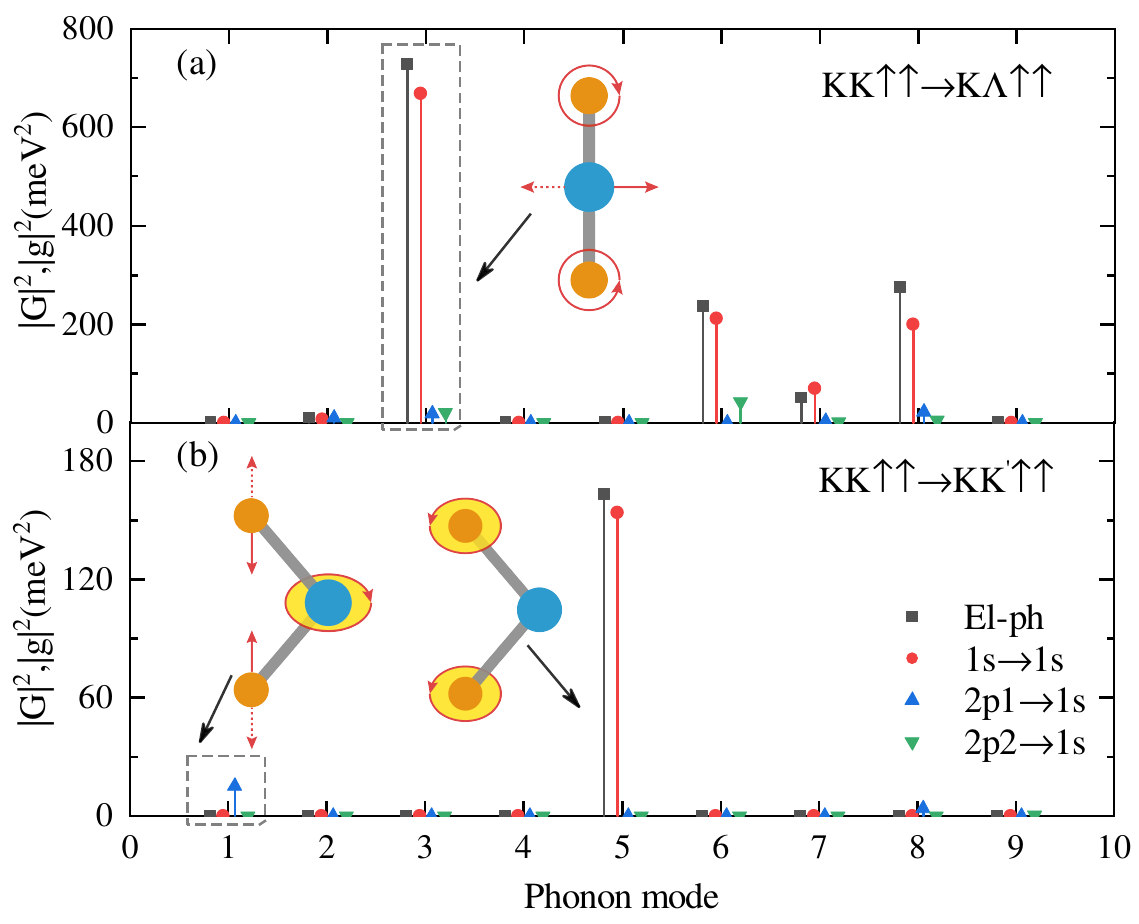}
	\caption{The modulus square of the ExPC matrix elements, $|G|^{2}$, for the scattering from three initial states, KK$\uparrow \uparrow$ 1$s$ (red circle), KK$\uparrow \uparrow$ 2$p$1 (blue triangle), KK$\uparrow \uparrow$ 2$p$2 (green inverted triangle) to two final states, K$\Lambda$$\uparrow\uparrow$ 1$s$ (a), KK$^{\prime}$$\uparrow \uparrow$ 1$s$ (b).
	The modulus squares of the EPC matrix elements ($|g|^{2}$) from K$\uparrow$$\rightarrow$$\Lambda$$\uparrow$ (a) and K$\uparrow$$\rightarrow$K$^{\prime}$$\uparrow$ (b) are also shown by black solid squares for comparison. Phonon modes 1 to 9 are numbered in ascending order in energy. In the side-view illustration of atomic vibrations (inset), an ellipse denotes the in-plane (of the 2D layer) rotation of phonon eigenvector and a circle denotes the out-of-plane rotation of phonon eigenvector.
	}
	\label{fig2}
\end{figure}
\begin{figure*}[t]
	\centering
	\includegraphics[width=0.9\linewidth]{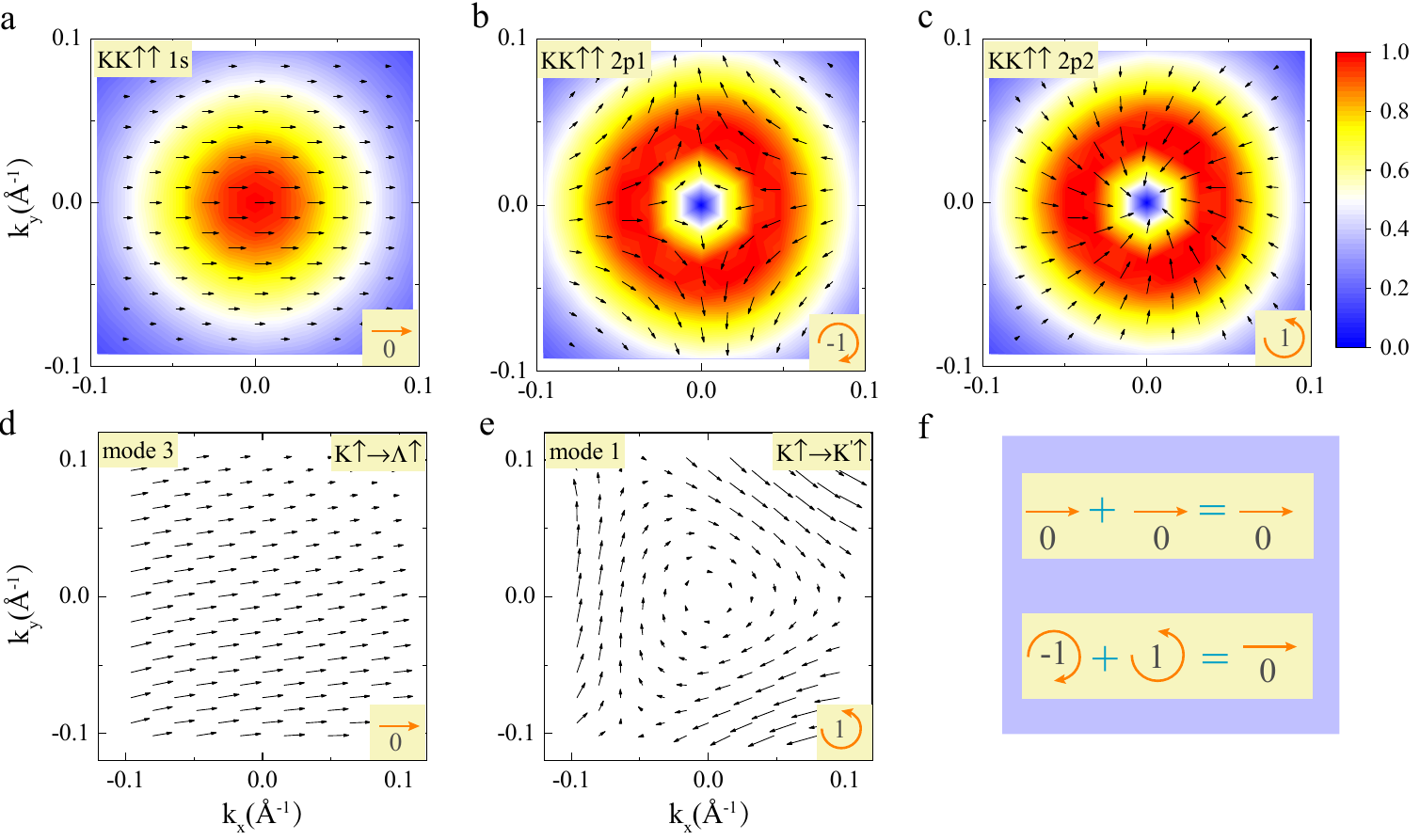}
	\caption{(a)-(c) The distribution of the calculated exciton envelope functions of 1$s$, 2$p$1 and 2$p$2 states of the KK $\uparrow \uparrow$ excitons in the BZ. The envelope functions are normalized by their largest values. The color and the length of an arrow denote the distribution weight. The direction of an arrow denotes the phase of envelope function.
		(d)-(e) The distribution of the calculated EPC matrix elements in the BZ for the electron-phonon scattering of K$\uparrow$$\rightarrow$$\Lambda$$\uparrow$ and K$\uparrow$$\rightarrow$ K$^{\prime}$$\uparrow$ by different phonon modes. The length and the direction of an arrow denote the magnitude and the phase of an EPC matrix element, respectively. (d)-(e) are used to calculate the exciton-phonon scattering from three KK$\uparrow \uparrow$ states to two final states, i.e. K$\Lambda$$\uparrow\uparrow$ 1$s$ and KK$^{\prime}$$\uparrow \uparrow$ 1$s$, respectively.
		The origin is located at the extreme $k$-point K. The orange arrow in the corner denotes the topology of the distribution of envelope functions and EPC matrix elements around the extreme k-point. And the numbers in the corner denotes the exciton angular quantum number or winding number of EPC matrix elements. (f) The winding number of EPC matrix elements contributes to the change of exciton angular quantum number for the allowed scattering channel.
	}
	\label{fig3}
\end{figure*}

First-principles calculations of the ExPC matrix elements $G$ in Eq.~\ref{ex-ph-G} are numerically intensive and challenging.
First, accurate descriptions of the exciton envelope function of the BSE requires very dense Brillouin zone (BZ)-sampling, especially in 2D semiconductors~\cite{qiu2013optical,qiu2016screening}.
Finite-momentum excitons involved in the exciton-phonon problems further increases the complexity of the problem.
Second, the gauge degree of freedom in the calculations of the exciton envelope function and the EPC matrix elements often causes phase mismatch between the two.
Owing to these difficulties, accurate first-principles calculations of ExPC matrix elements and scattering times for noncentrosymmetric monolayer semiconductors remain challenging~\cite{reichardt2020nonadiabatic,huang2021exciton}.
Here we employ a newly-developed algorithm to calculate ExPC matrix element $G$ in WSe$_2$, a typical monolayer transition metal dichalcogenide (TMD), to investigate the exciton-phonon scattering of 2D semiconductors.
This system is chosen due to its large excitonic effects and emergence of multiple nearly degenerate valleys in the BZ, as discovered in recent experiments~\cite{he2014tightly}.
In Eq.~\ref{ex-ph-G}, exciton envelope functions and EPC matrix elements are calculated by first-principles GW-BSE method implemented in BerkeleyGW~\cite{rohlfing2000electron} and density-functional perturbation theory (DFPT) implemented in Quantum ESPRESSO~\cite{giannozzi2009quantum}, respectively.
As mentioned before, although the ExPC matrix element $G$ is gauge-invariant (up to a trivial phase),
the exciton envelope functions and EPC matrix elements that give rise to $G$ in Eq.~\ref{ex-ph-G} can carry arbitrary gauges and introduce numerical difficulties in first-principles calculations as well as the interpretations of results.
In bulk materials, this gauge problem can be solved by using the same uniform $k$-grid which carries identical gauge for both GW-BSE and EPC calculations in the ExPC problem~\cite{chen2020exciton}.
However, in 2D materials, special attention should be paid because very dense $k$-point sampling is needed to converge the exciton calculations, making the usage of a uniform $k$-grid across the whole BZ in both the GW-BSE and DFPT calculations computationally challenging.
In monolayer WSe$_2$, we find that the $k$-grid density equivalent to 14,400 points in the BZ is necessary to capture the exciton energy and resolve the exciton wave functions (see Figs.~\ref{fig3}(a)-(c)), consistent with literature results~\cite{man2021experimental}.
To resolve this issue, we use the patch-sampling method in GW-BSE, together with gauge-fixing techniques applied to both the exciton envelope function and the EPC matrix elements.
A physically motivated hydrogen gauge has been applied, making the envelope function of the 1$s$ exciton real~\cite{cao2018unifying}.
This hydrogen gauge allows us to denote the exciton bands using the principle and the angular quantum numbers of 2D Rydberg atoms, and define a winding number of the EPC matrix elements for the electron-phonon selection rule in Figs.~\ref{fig3}(d)-(e).
Alternatively, one could set the gauge by making the largest Fourier component of the Bloch function real~\cite{giustino2007electron}.
However, under such a gauge, the exciton envelope function of the 2D Wannier excitons would not map to the wavefunction of 2D hydrogen atoms, which obscures the exciton-phonon selection rule that we will discuss later.
Other calculation details are included in Sec. I-II of the Supplemental Materials.
Figs.~\ref{fig1}(a)-(b) schematically show the phonon-mediated relaxation paths of the direct exciton in monolayer WSe$_2$.
Possible initial states considered in our study are 1$s$ and 2$p$ direct excitons, denoted as KK$\uparrow\uparrow$ 1$s$ and KK$\uparrow\uparrow$ 2$p$ excitons.
Here we label the excitons by the valley-spin configuration of their constituent electron and hole, with that of the hole on the left.
A KK$\uparrow\uparrow$ 1$s$ exciton can be resonantly excited by a circularly-polarized
photon~\cite{cao2012valley,xiao2012coupled}.
After absorbing two (or more) circularly-polarized photons, a KK$\uparrow\uparrow$ 2$p$ exciton can be formed, which may further relax into lower-energy and/or indirect excitons.
Figs.~\ref{fig1}(d)-(f) show eigenvalues of the direct and indirect 1$s$ and 2$p$ excitons from GW-BSE calculations.
We note that in our calculations, the energies of K$\Lambda\uparrow\uparrow$ and K$\Lambda^{\prime}\uparrow\downarrow$ 1$s$ excitons are smaller than KK$\uparrow\uparrow$ and KK$\uparrow\downarrow$ 1$s$ excitons by 0.1-0.2 eV, which is slightly different from the approximate degeneracy found in a recent work~\cite{madeo2020directly}.
This discrepancy arises from the choice of exchange-correlation functional and lattice constants used in the mean-field calculation (see Fig.~S3), and as we will show later, this minor difference does not affect the analysis of ExPC.
To unveil the rules of exciton-phonon scattering and how they differ from the electron-phonon scattering, we select two final exciton states, i.e.,
K$\Lambda\uparrow\uparrow$ 1$s$ and KK$^{\prime} \uparrow \uparrow$ 1$s$, in the analysis.
Here we focus on the ExPC matrix elements in Eq.~\ref{ex-ph-G}, and extract the essential physics of scattering between initial and final exciton bands.
In the subsequent calculations of scattering time, we include the energy conservation relation in the exciton-phonon scattering events to account for the density-of-state effects.
We note that, during the exciton-phonon scattering events of interest, while the electron is scattered to a different valley, the hole in monolayer WSe$_2$ resides in the same K valley.
As a result, only the first term (electron-phonon scattering) of $G$ in Eq.~\ref{ex-ph-G} can be non-zero.
Phonon-mode dependent EPC and ExPC matrix elements are shown in Fig.~\ref{fig2}.
At the electron-phonon level, the matrix elements corresponding to inter-valley scatterings of electrons from K to K$^{\prime}$ or $\Lambda$.
The significant phonon-mode dependence of the EPC matrix elements can be understood by group theory analysis (details in Sec.~II of the Supplemental Materials).
The groups of wavevectors are $C_{1h}$ and $C_{3h}$ at $\Lambda$($\Lambda^{\prime}$) and K (K$^{\prime}$), respectively, making the electron-phonon selection rules accessible by transforming the EPC matrix elements under an in-plane reflection $\hat{\sigma}_{h}$ and/or three-fold rotation $\hat{C}_{3}$.
For K$\uparrow\rightarrow\Lambda\uparrow$ scattering, the allowed phonon modes, i.e. the second, third, sixth, seventh, and eighth modes in Fig.~\ref{fig2}(a), are symmetric under $\hat{\sigma}_{h}$.
For K$\uparrow\rightarrow$K$^{\prime}\uparrow$, the allowed phonon modes need not only to be symmetric under $\hat{\sigma}_{h}$, but also to conserve the three-fold pseudo-angular momentum~\cite{zhang2015chiral}.
As a result, only the fifth mode is allowed as shown in Fig.~\ref{fig2}(b).

The matrix elements calculated at the exciton-phonon levels exhibit distinct features from those at the electron-phonon level due to the influence of exciton envelope function that carries angular momentum (Eq.~\ref{ex-ph-G}).
First, excitons with different principle and angular quantum numbers interact differently with a given phonon mode, a feature missing in the electron-phonon level.
For example, for the third mode in Fig.~\ref{fig2}(a), although the scattering strength of 1$s$ $\rightarrow$1$s$ is similar to that of electron-phonon scattering, the strengths of 2$p$1$\rightarrow$1$s$ and 2$p$2$\rightarrow$1$s$ decrease sharply.
More interestingly, conventional group theory analysis of the EPC matrix element show that the first mode in Fig.~\ref{fig2}(b) cannot contribute to the scattering.
In the exciton-phonon level, however, the strength of 2$p$1$\rightarrow$1$s$ is greatly enhanced, exceeding the strength of the 1$s$ $\rightarrow$1$s$ process.
Similar distinctions between the ExPC and EPC matrix elements have also been found for the scattering into K$\Lambda^{\prime}$ exciton (Fig.~S5).
These unusual exciton-phonon scattering patterns motivate us to explore the selection rules generic to the exciton-phonon scattering problems in a 2D Wannier exciton system.
Since the envelope function of the first few low-energy Wannier exciton is highly localized in $k$-space, in Eq.~\ref{ex-ph-G}, the summation over $k$-points can be performed in the vicinity of the respective band extreme of the hole at $\bm{k}_{0}$, or the electron at $\bm{k}_{0}+\bm{Q}$.
The envelope function of the exciton Rydberg states can be written as~\cite{cohen2016fundamentals}:
\begin{equation}
A_{v\bm{k},c\bm{k}+\bm{Q}}^{S}\approx R_{nl} (\lvert \bm{k^{\prime}} \rvert) e^{il\theta_{\bm{k}^\prime}},
\label{A-atom}
\end{equation}
where $\bm{k}^{\prime}=\bm{k}-\bm{k}_{0}$. $\theta_{\bm{k}^\prime}$ denotes the azimuthal angle of $\bm{k}^\prime$. $n$ and $l$ are the principal and angular quantum numbers of exciton $S$.
The radial part, $R_{nl} (\lvert \bm{k^{\prime}} \rvert)$, is real while the angular part, $e^{il\theta_{\bm{k}^{\prime}}}$, rotates in the complex plane for non-zero $l$.
For a particular phonon mode $\nu$ that scatters electrons between two non-degenerate bands, the continuity of $k$-resolved EPC matrix elements allows us to map them to a vector field tangential to the BZ, and use the winding number to character the phase of the EPC matrix elements.
For example, to scatter an electron near the conduction band minimum at $\bm{k_0}+\bm{Q}$, the EPC matrix elements can be written as
\begin{align}
    g_{c'c}(\bm{k}+\bm{Q},\bm{q}\nu)\approx T_{c'c,m}(|\bm{k}'|,\bm{q}\nu)e^{im\theta_{\bm{k}'}},
    \label{g-wind}
\end{align}
where $m$ is the winding number.
In the case of exciton-phonon scattering that changes the momenta of bound electrons (e.g., monolayer WSe$_2$), the ExPC matrix element (Eq.~\ref{ex-ph-G}) can be rewritten as:
\begin{align}
	G_{n'l',nl}(\bm{Q},\bm{q}\nu)&\approx \sum_{\bm{k}'}\Tilde{R}_{n'l'}(|\bm{k}'|)R_{nl}(|\bm{k}'|) T_{c'c,m}(|\bm{k}'|,\bm{q}\nu)\nonumber \\
&\qquad \qquad \qquad \qquad \times e^{i(l-l'+m)\theta_{\bm{k}'}}
	\label{G-atom}.
\end{align}
Eq.~\ref{G-atom} involves the dominant contribution from the first term in Eq.~\ref{ex-ph-G}.
For simplicity, we include one valence and one conduction band.
The generalization to hole-phonon scattering and multi-band case is straightforward.
From Eq.~\ref{G-atom}, the winding number of EPC matrix elements should contribute to the change of exciton angular quantum numbers for the allowed scattering, giving rise to the selection rule,
\begin{align}
    l - l' + m = 0
	\label{Selection-rule}.
\end{align}
Eq.~\ref{Selection-rule} shows that ExPC is fundamentally different from EPC, and gives unexpected phonon-mediated scattering between excitons of different angular momentum if the EPC winding number $m$ is non-zero.
The selection rules in Eq.~\ref{Selection-rule} can be used to understand the results from first-principles calculations in Fig.~\ref{fig2}.
Figs.~\ref{fig3}(a)-(c) show normalized envelope functions of 1$s$, 2$p$1 and 2$p$2 states of the KK $\uparrow \uparrow$ exciton in the BZ.
They are all very localized with the extension radius of $\sim$0.1~$\text{\AA}^{-1}$, highlighting the importance of dense k-sampling for GW-BSE calculations.
The quantum phase of envelope functions allows us to identify 2$p$1 and 2$p$2 states that have angular quantum numbers $l$ of -1 and 1, respectively.
The two final 1$s$ states, K$\Lambda\uparrow\uparrow$ and KK$^{\prime}\uparrow\uparrow$, both have $l^{\prime} = 0$.
Figs.~\ref{fig3}(d)-(e) show the $k$-resolved EPC matrix elements between different electron valleys, for processes highlighted in Fig.~\ref{fig2}.
The K$\rightarrow\Lambda$ ($\nu = 3$) and K$\rightarrow K^{\prime}$ ($\nu = 1$) EPC matrix elements have winding numbers $m$ of 0 and 1, respectively.

\begin{figure}[b]
	\centering
	\includegraphics[width=1.0\columnwidth]{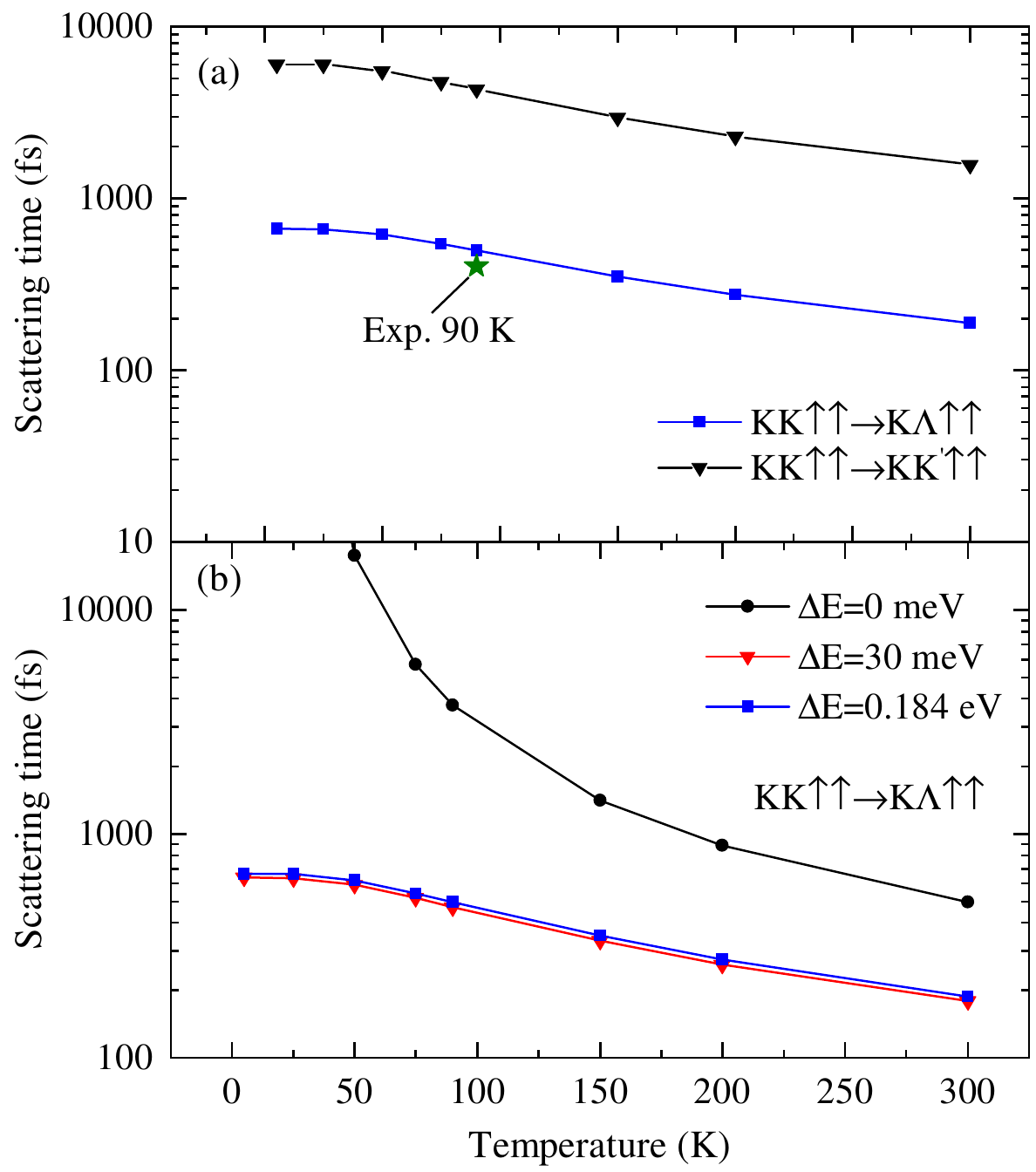}
	\caption{(a) Calculated temperature-dependent scattering times of KK 1$s$ excitons for two channels. The experimental value from Ref.~\cite{madeo2020directly} at 90 K is marked by a star. (b)The temperature-dependent scattering times for different energy differences between initial and final states. In (a), the energy difference $\Delta$E used is 0.18 eV.
	}
	\label{fig4}
\end{figure}

We next perform symmetry analysis to understand why non-zero winding numbers in the EPC matrix elements could occur in Fig.~\ref{fig3}(e).
For these intervalley K$\uparrow\rightarrow$K$^{\prime}$$\uparrow$ processes connecting the two high-symmetry points, the scattering mediated by the first phonon mode is forbidden due to the EPC selection rules.
However, the scattering becomes allowed at neighboring k-points of $\bm{K}$ due to symmetry reduction.
Therefore, geometrically, $\bm{K}$ becomes a critical point in the vector-field representation of EPC matrix elements in Fig.~\ref{fig3}(e), leading to a winding number of 1.
According to Eq.~\ref{Selection-rule}, this winding number 1 could compensate the angular quantum number -1 of the KK$\uparrow \uparrow$ 2$p$1 exciton, and allow the first phonon mode to scatter KK$\uparrow \uparrow$ 2$p$1 $\rightarrow$ KK$^{\prime} \uparrow \uparrow$ 1$s$ (Fig.~\ref{fig2}(b)).
%
%In other words, as shown in %Fig.~\ref{fig3}(f), the new selection %rules implied in Fig.~\ref{fig2} can be
%well describted.
%the compensation
%of the angular quantum numbers of the %exciton envelope function and the winding %numbers of the EPC matrix elements.
%
This is in sharp contrast in Fig.~\ref{fig3}(d).
Since $\bm{\Lambda}$ has the same symmetry as that of its neighboring k-points, the K$\uparrow\rightarrow\Lambda\uparrow$ ($\nu$=3) EPC matrix elements are nearly the same.
In this case, the continuous and finite EPC matrix elements have a winding number of 0.
As a result, the third mode could contributes to the scattering KK$\uparrow \uparrow$ 1$s$ $\rightarrow$ K$\Lambda\uparrow\uparrow$ 1$s$ rather than K$\Lambda\uparrow\uparrow$ 2$p$1 or K$\Lambda\uparrow\uparrow$ 2$p$2.
Finally, we apply our first-principles algorithm and selection rules to study the phonon-mediated inter-valley relaxation pathways of KK$\uparrow\uparrow$ 1$s$ excitons, and validate our theory using a recent experimental measurement.
Phonon modes which satisfy the energy conservation relation are sampled in the phonon BZ, where a 30×30×1 $q$-grid is used and interpolated into 240×240×1.
Other calculation details are included in Sec.~IV of the Supplemental Materials.
Fig.~\ref{fig4}(a) shows the temperature-dependent scattering times of the two dominant processes, i.e., KK$\uparrow \uparrow$$\rightarrow$K$\Lambda$$\uparrow \uparrow$ and KK$\uparrow \uparrow$$\rightarrow$KK$^{\prime}$$\uparrow \uparrow$.
The temperature dependence is mainly caused by the phonon occupation number.
At 90~K, the calculated scattering time of KK$\uparrow\uparrow$$\rightarrow$K$\Lambda$$\uparrow \downarrow$ is about 500~fs, which is in good agreement with the measured value, $\sim$400~fs.
To show that the calculated scattering time is insensitive to the energy differences between the KK$\uparrow\uparrow$ 1$s$ and K$\Lambda$$\uparrow\uparrow$ 1$s$ excitons, we plot in Fig.\ref{fig4}(b) the temperature dependent scattering times for different energy differences.
In the degenerate case, We notice a divergent scattering time at low temperatures, because only phonon-absorption scattering process can take place and the phonon occupation number is near zero.
However, if KK$\uparrow\uparrow$ 1$s$ is higher than K$\Lambda\uparrow\uparrow$ 1$s$ by
30 meV (within the experimental error), the scattering time significantly decreases and becomes almost the same as $\Delta$E=0.18 eV, the value we used in Fig.\ref{fig4}(a).
Therefore, we conclude that the exciton intervalley scattering observed in Ref.~\cite{madeo2020directly} mainly arises from phonon-mediated
KK$\uparrow \uparrow$$\rightarrow$K$\Lambda$$\uparrow\uparrow$ process.
These numerical results shed light on the microscopic mechanism of exciton relaxations in experiments.
Considering the recent serge of research about excitonic properties in layered materials, especially TMD,
we believe the selection rules discovered and the computational method developed here can stimulate both theoretical and experimental studies of exciton-phonon interactions in 2D semiconductors.
Further studies that include spin-flip scattering channels may also unveil rich patterns in the EPC matrix element and unexpected scattering pathways.

\begin{acknowledgments}
X.Z., E.W., and X.L. are supported by the National Science Foundation of China under Grant
Nos. 11934003, 11774003, and 11634001, the Beijing Natural Science Foundation under Grand No. Z200004, and
the Strategic Priority Research Program of the Chinese Academy of Sciences under Grant No. XDB33010400 in first-principles calculations.
K.X. and T.C. are supported by NSF through the University of Washington Materials Research Science and Engineering Center Grant No. DMR-1719797 in theoretical analysis.
The computational resources were supported by the high-performance computing platform of Peking University, China.
\end{acknowledgments}

\end{document}